\documentclass[twocolumn,prb,showpacs,floatfix,superscriptaddress]{revtex4}%
\usepackage{graphicx}
\usepackage{amsmath}
\usepackage{amsfonts}
\usepackage{xcolor}
\usepackage{amssymb}%
\newcommand{\lirho}{Li$_2$RhO$_3$}
\newcommand{\liiro}{Li$_2$IrO$_3$}
\newcommand{\nairo}{Na$_2$IrO$_3$}
\providecommand{\U}[1]{\protect\rule{.1in}{.1in}}
\begin{document}
\title{Origin of the insulating state in honeycomb iridates and rhodates }
\author{I. I. Mazin}
\affiliation{Code 6393, Naval Research Laboratory, Washington, DC 20375, USA}
\author{S. Manni}
\affiliation{I. Physikalisches Institut, Georg-August-Universit\"at G\"ottingen, 37077 G\"ottingen, Germany}
\author{K. Foyevtsova$^*$}
\affiliation{Institut f\"ur Theoretische Physik, Goethe-Universit\"at Frankfurt, 60438 Frankfurt am Main, Germany}
\author{Harald O. Jeschke}
\affiliation{Institut f\"ur Theoretische Physik, Goethe-Universit\"at Frankfurt, 60438 Frankfurt am Main, Germany}
\author{P. Gegenwart}
\affiliation{I. Physikalisches Institut, Georg-August-Universit\"at G\"ottingen, 37077 G\"ottingen, Germany}
\author{Roser Valent{\'\i}}
\affiliation{Institut f\"ur Theoretische Physik, Goethe-Universit\"at Frankfurt, 60438 Frankfurt am Main, Germany}
\date{\today }


\begin{abstract}
  A burning question in the emerging field of spin-orbit driven
  insulating iridates, such as {\nairo} and {\liiro} is whether the
  observed insulating state should be classified as a Mott-Hubbard
  insulator derived from a half-filled relativistic $j_{\rm eff}=1/2$
  band or as a band insulator where the gap is assisted by spin-orbit
  interaction, or Coulomb correlations, or both.  The difference
  between these two interpretations is that only for the former
  strong spin-orbit coupling ($\lambda\gtrsim W,$ where $W$ is the
  band width) is essential.  We have synthesized the isostructural and
  isoelectronic {\lirho} and report its electrical resistivity and
  magnetic susceptibility. Remarkably it shows insulating behavior
  together with fluctuating effective $S=1/2$ moments, similar to
  {\nairo} and {\liiro}, although in Rh$^{4+}$ (4$d^{5}$) the
  spin-orbit coupling is greatly reduced. We show that this behavior
  has non-relativistic one-electron origin (although Coulomb
  correlations assist in opening the gap), and can be traced down to
  formation of quasi-molecular orbitals, similar to those in {\nairo.}
\end{abstract}
\maketitle

In recent years complex iridium oxides have caused extraordinary
interest~\cite{Moon,Okamoto,Kim,Pesin} since the physics there is
governed by a unique combination of several comparable scales:
one-electron hopping $t,$ spin-orbit coupling (SOC) $\lambda,$ and the
Hubbard repulsion $U$.  The honeycomb layered compound {\nairo}, a
small-gap antiferromagnetic insulator~\cite{Singh10,Comin12}, has
received particular attention. It was suggested that the adequate
description of the electronic behavior of this system is in terms of a
half-filled relativistic $j_{\rm eff}=1/2$ band, which becomes a
Mott-Hubbard insulator~\cite{Chaloupka10}. However, $U$ in iridates is
rather small (1.5--2 eV), therefore the corresponding band must be
rather narrow for the system to become insulating.  In this scenario,
the SOC is the leading interaction in these systems, so that the
$t_{2g}$ bands split into a narrow doublet with the effective angular
moment $j_{\rm eff}=1/2$ and a quartet with $j_{\rm eff}=3/2.$ 
In the idealized crystal structure, the 
one-electron hopping between the doublet states is fully suppressed,
and the effect of one electron hopping is reduced, by perturbation theory, 
to the second order in $t$, that is, to $t^2/1.5\lambda\sim t/3$,
where 1.5$\lambda$ is the energy separation between the doublet and
the quartet.

Recently, some of us~\cite{Mazin12} proposed an alternative description
and argued that the one-electron non-relativistic band structure might
be a better starting point for the description of the electronic
behavior of honeycomb iridates than the limit $\lambda\gg W$ (band
width). In this case the band structure is dominated by the formation
of so-called quasi-molecular orbitals (QMOs), and consists of four
narrow bands (the width being defined by second-neighbor hoppings and
other secondary one-electron parameters), spread over a width of
$\sim4t,$ where $t\sim0.3$ eV is the leading one-electron hopping. The
highest and the lowest bands are singlets, having one state per two Ir
($i.e.,$ one state per spin per unit cell of two formula units), and
the two middle bands are doublets. In the $\lambda=0$ limit the upper
singlet and doublet bands nearly merge, forming a triplet manifold,
while turning on SOC further splits those bands into three
singlets. The upper two bands barely overlap, forming an incipient
(SOC assisted) band insulator, and even a very small $U$ of a few
tenth of an eV is sufficient to open a gap. In this picture, the
material is characterized as a spin-orbit assisted insulator with the
gap enhanced by Hubbard correlation.

In view of the two alternative, and partly opposite, descriptions of
the insulating state in these systems, we present here a comparative
analysis between the electronic behavior of hexagonal iridates and
rhodates.  Specifically, we have synthesized and investigated
{\lirho}, which shows insulating behavior at low temperatures, similar
to {\nairo} and {\liiro}, even though in Rh$^{4+}$ (4$d^{5}$) the SOC
is substantially reduced. The comparison sheds light onto the nature
of the insulating state in these systems.

The paper is organized as follows. We first settle the terminology
between the various definitions of insulators. We then proceed with
electrical resistivity and magnetic susceptibility data of {\lirho}
and the description of its electronic and magnetic properties by means
of density functional theory (DFT) calculations with and without
inclusion of spin-orbit coupling and discuss the similarities and
differences of this rhodate system compared to the hexagonal
iridates. As an outlook, we provide some predictions for the magnetism
in the hexagonal rhodates.

The question of whether a particular phase is characterized as a
Mott-Hubbard insulator or a band insulator is largely terminological,
as no strict definition or criterion exists to rigorously separate
these notions.  Some authors~\cite{Gebhard} further subdivide the
classification of insulators into Peierls, Wilson, Slater or Hund
insulators, to mention a few.  We feel that this fine tuning is not
helpful here, and will concentrate on the difference between Mott and
band insulators, which is fundamental in the sense that one cannot go
from the former to the latter continuously. Note that this division
does not have a one-to-one correspondence with the strongly correlated
-- weakly correlated dichotomy; a band insulator, in our terminology,
may have a gap strongly enhanced by correlations, but is still
\textquotedblleft topologically connected\textquotedblright\ with an
uncorrelated insulator.

We can illustrate this with a simple example: imagine a crystal of
atoms with one half-occupied orbital. If the crystal has one atom per
unit cell, then on the one-electron level this material can never be
insulating. Upon including an onsite Hubbard $U$, thereby penalizing double
occupation and hindering itinerancy, it becomes a {\it Mott
  insulator}~\cite{GeorgeRMP,Imada1998}, with no coherent
quasiparticles. This happens roughly when $U$ becomes larger than the
total band width $W.$ Now, suppose the same atoms are bound in dimers
forming a molecular crystal. Each dimer develops a bonding and an
antibonding state split by some energy $\Delta.$ We now allow
inter-dimer hopping. The levels will broaden into bonding and
antibonding bands of the width $W.$ As long as $W<\Delta,$ the
material is a {\it band insulator}. If $\Delta\sim W,$ the gap is very
small, and the indirect gap may even become negative. If we add a
Hubbard $U$ to this system (not necessarily larger than $W)$ the gap
will get larger by some fraction of $U$ (depending on the degree of
itinerancy) and this may be a substantial enhancement.  We call this a
{\it correlation-enhanced band insulator}.  For instance, solid Ne is
a band insulator, even though in LDA calculations its gap is severely
underestimated (12.7 eV {\it vs}. 21.4 eV).~\cite{Ne} This discrepancy
is related to another problem in the density functional theory (DFT),
the so-called density derivative discontinuity, and not to Hubbard
correlations.  To first approximation, this discrepancy is
inversely proportional to the static dielectric function.~\cite{gap}

An example of a Mott insulator is FeO. It has one electron in the
spin-minority $t_{2g}$ band, and is a metal in DFT. Coulomb
correlations have to destroy entirely the coherent DFT metallic band
crossing the Fermi level, and the excitation gap appears between the
incoherent lower and upper Hubbard band.~\cite{feo} Note that despite
FeO being a Mott insulator even in the paramagnetic phase, the
simplistic treatment of LDA+U,~\cite{MA} as opposed to more
sophisticated DMFT,~\cite{feo} cannot reproduce insulating behavior by
symmetry; the cubic symmetry needs to be broken, for instance by
assuming antiferromagnetic ordering, after which a gap opens at
sufficiently large values of $U$.  Similarly, in parent compounds of
the superconducting cuprates there exists one hole in the $e_{g}$
band, and symmetry does not allow to open a gap in DFT. These systems
are ``true'' Mott-Hubbard insulators.

Finally, MnO is an example of a (strongly) correlation-enhanced band
insulator. It has a gap between $3d$ majority and minority bands. In
DFT, this gap is driven by the Hund's rule, and is $\sim 5I-W,$ where
the Stoner factor $I$ is $\sim 1$ eV.  The calculated value is 1.4 eV
as compared to 4.5 eV in the experiment.~\cite{MnO} This material is
strongly affected by Mott physics, and routinely called Mott
insulator, yet one can make a \textit{gedanken experiment} and
gradually reduce the Hubbard correlations to zero, whereupon the gap
will drop to its DFT value, without losing the insulating character.
Note also that a Mott-Hubbard insulator, in our nomenclature, does not
necessarily imply a strong Hubbard repulsion $U\gg t,$ where $t$ is a
typical intersite hopping. For instance, TaS$_{2}$ by no means can be
expected to be a strongly correlated material, and $U$ cannot be more
than a fraction of an eV, and, indeed, at high temperatures it is a
metal. Yet at low temperature it experiences a charge density wave
transition typical for this structure, which, combined with the
spin-orbit interaction on Ta, splits off, essentially accidentally, an
ultra-narrow band ($W\sim0.1$ eV), and even a minuscule $U$ suffices
to split it into two Hubbard bands.

In {\nairo}, $U$ is relatively small and the material cannot be
strongly correlated. Moreover, correlations are suppressed in LDA+U
calculations because of substantial delocalization of electrons over
Ir$_{6}$ hexagons, so that in order to increase the calculated gap
from $\sim0$ to $\sim0.3$ eV one has to add $U\sim4$ eV. Furthermore,
there is indirect evidence of itinerancy in the experiment: the
ordered magnetic moment even at the lowest temperature is less than
$0.3\mu_{B}$,~\cite{Radu} in reasonable agreement with the band
structure calculations ($\approx0.5$ $\mu_{B},$ equally distributed
between spin and orbital moments), while the fully localized $j_{\rm
  eff}$ model ($1\mu_{B},$ split 2:1 between spin and orbital moments)
requires strong fluctuations to suppress the ordered
moment.~\cite{Chaloupka13}

It is often argued that the experimentally measured~\cite{Clancy12}
spin-orbit correlation factor, $\left\langle \mathbf{L\cdot
    S}\right\rangle ,$ is consistent with one hole in the $j_{\rm
  eff}=1/2$ state, and thus proves its existence.  However, this
factor is mostly collected from the $e_{g}$
holes,~\cite{Haskel,Foyevtsova13} and is well described by band
structure calculations.

A comparison of honeycomb iridates with the isostructural and
isoelectronic {\lirho} should be very instructive, because if the
former are SOC Mott insulators like Sr$_2$IrO$_4$,~\cite{Kim,Li2013}
then a Rh analogue (with much weaker SOC) should be metallic, just as
Sr$_{2}$RhO$_{4}$.~\cite{Martins11} If, on the other hand, the
formation of quasi-molecular orbitals triggers the insulating
behavior, 
then a larger $U$ in Li$_{2}$RhO$_{3}$ will likely recreate the same
physics as 
for {\nairo}, i.e. a correlation-enhanced band insulator.

\begin{figure}[ptb]
\begin{center}
\includegraphics[width=\columnwidth]{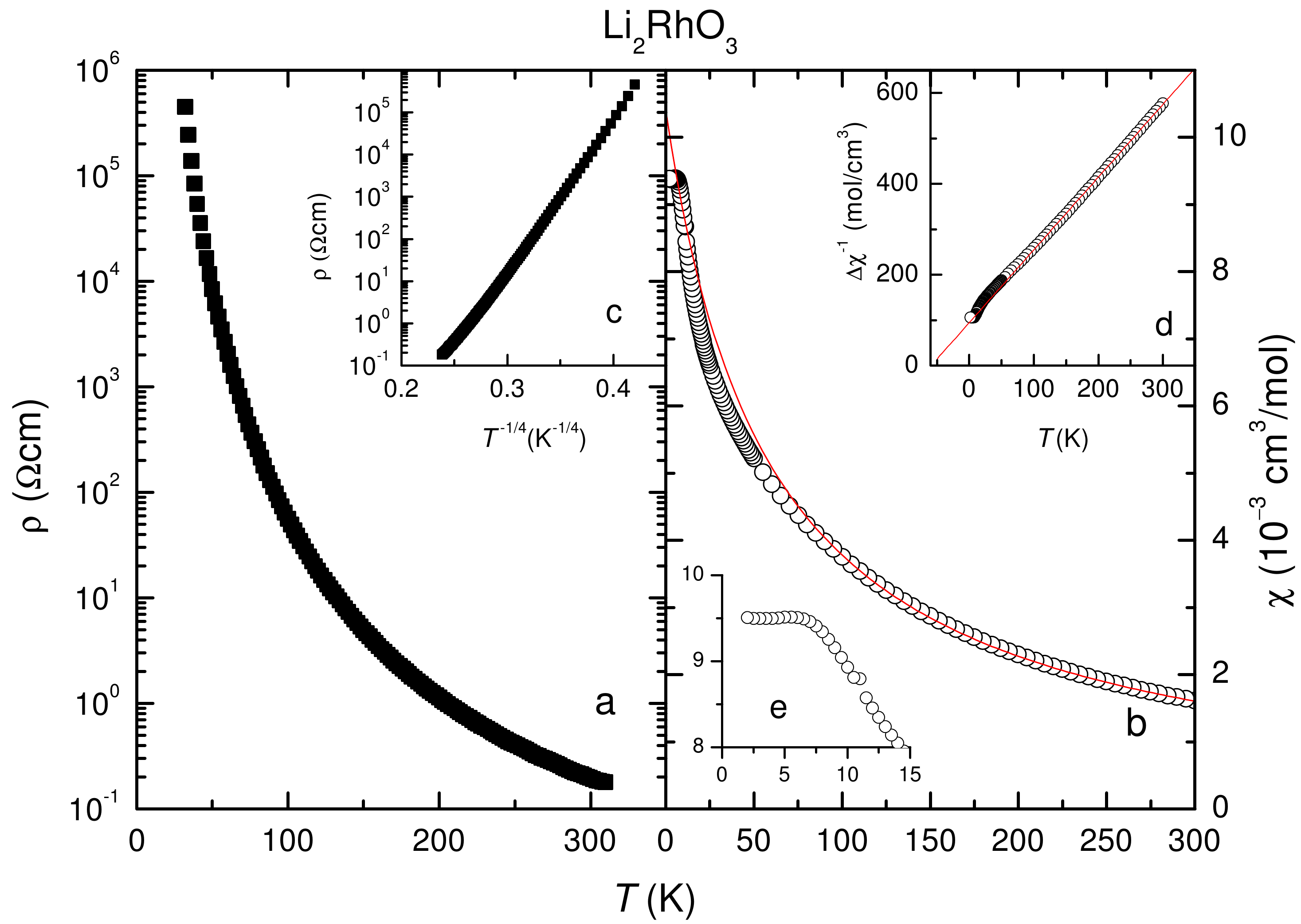}
\end{center}
\caption{(Color online) Temperature dependence of the electrical
  resistivity (a) and magnetic susceptibility (b) of polycrystalline
  Li$_2$RhO$_3$.  Inset (c) displays the resistivity data (on log
  scale) vs. $T^{-1/4}$. The line in (b) is a Curie-Weiss (CW) fit
  $\chi(T)=\chi_0+C/(T-\Theta_W$ with $\chi_0=-1.235\times 10^{-4}$
  cm$^3$/mol and $\Theta_W=-59$~K.  The inset (d) shows $\Delta
  \chi^{-1}$ vs $T$ with $\Delta\chi=\chi(T)-\chi_0$, and the CW
  fit. Inset (e) displays the same data as in (b) for the
  low-temperature regime.}
\label{exp}%
\end{figure}

\begin{figure}[ptb]
\includegraphics[width=.5\textwidth]{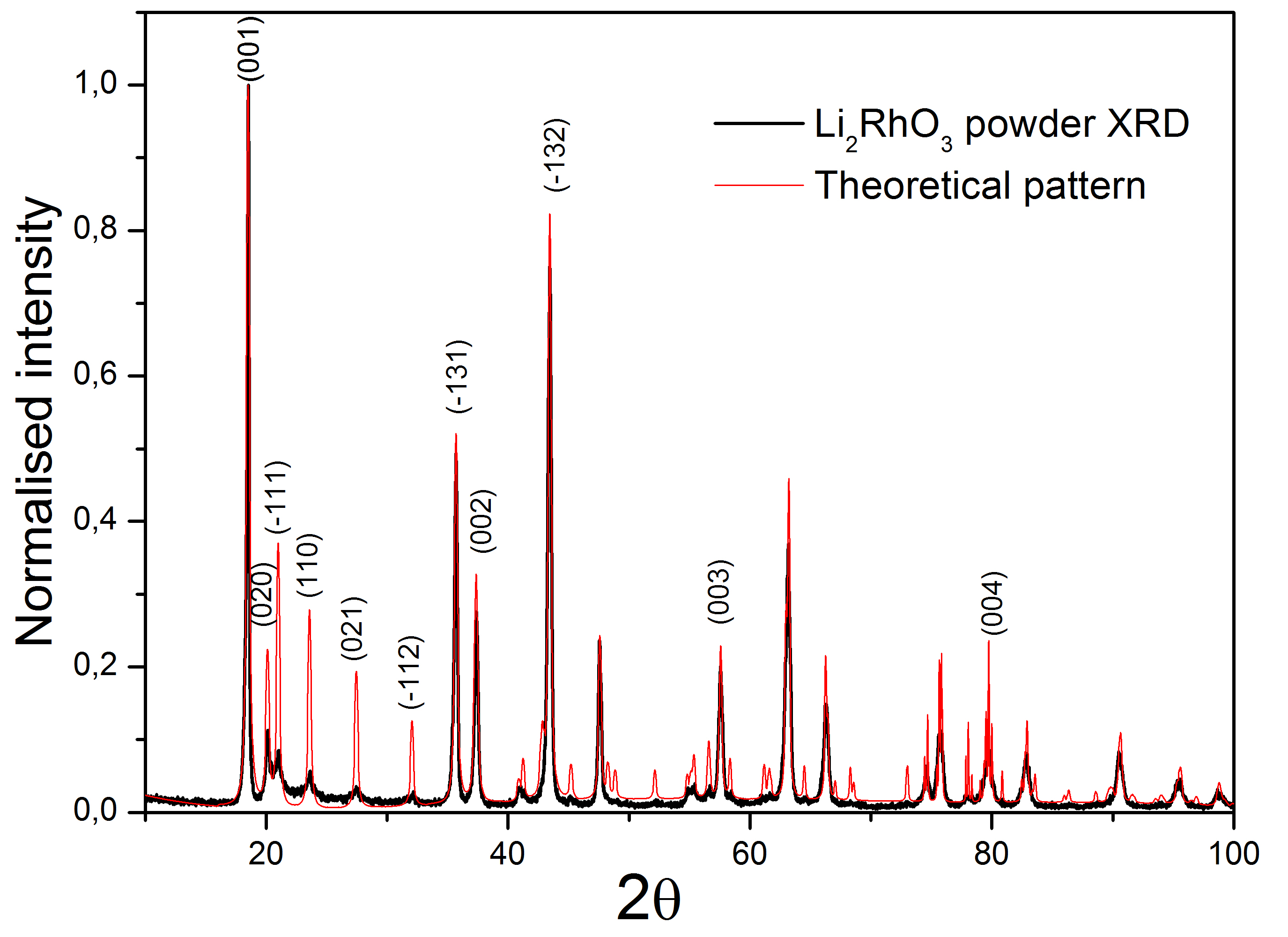}
\caption{(Color online) Comparison of observed (black) and calculated
  (red) powder XRD spectra for Li$_2$RhO$_3$ (see text).}
\label{powder}
\end{figure}

We have synthesized Li$_{2}$RhO$_{3}$ polycrystals by the solid state
reaction method from stoichiometric amounts of Li$_2$CO$_3$ and Rh
powder. The mixture has several times been pelletized and reacted in
O$_2$ flow at temperatures up to 850$^\circ$C. Powder X-ray
diffraction (XRD) scans do not reveal any evidence for secondary
phases and are similar to those reported in
Ref.~\onlinecite{Todorova}, see Fig.~\ref{powder}.  Magnetic
susceptibility and (four-probe) electrical resistivity have been
determined utilizing commercial (Quantum Design) instruments.

As presented in Fig.~\ref{exp}~(a), Li$_{2}$RhO$_{3}$ shows clear
insulating resistivity behavior, which follows the same variable-range
hopping dependence as found in {\nairo} or
{\liiro}.~\cite{Singh10,Singh12} Previous resistivity measurements at
higher temperatures found an activation gap of $\sim
80$~meV.~\cite{Todorova} The magnetic susceptibility
(Fig.~\ref{exp}~(b)) is Curie-Weiss (CW) like, with a small kink at
6~K, likely due to some spin-glass freezing, which needs to be
investigated in future measurements. The CW fit between 100 and 300 K
corresponds to $\mu_{\rm eff} = 2.2\mu_B$. Similar results have been
recently reported by Luo {\it et al.}~\cite{luo}

Now, we need to establish the crystal structure. Lab powder XRD (PWXD)
is in not very sensitive to the O-positions, which hinders structural
determination.  For instance, initial powder XRD refinement for {\nairo}
~\cite{Singh10} was unable to distinguish between $C\,2/c$ and
$C\,2/m$ structure but later measurements on a single crystal, showed
that $C\,2/m$ is the most stable crystal structure with well ordered
regular honeycomb planes.~\cite{Radu} Also for Li$_{2}$RhO$_{3}$ there
has been discussion about Li/Rh site exchange
~\cite{Todorova}. However, we have found that site exchange and
stacking faults have similar effects on powder XRD Rietveld
refinement. Thus, from the present data it is very hard to distinguish
between them. On the other hand, single crystal XRD on {\nairo} by
Choi et al. found evidence that stacking faults are the leading
defects rather than Na/Ir site exchange.~\cite{Radu}.

\begin{table}[ptb]
  \caption{Optimized crystal structure of {\lirho}, using experimental
    lattice parameters ($a=5.123$~{\AA}, $b=8.836$~{\AA}, $c=5.885$~{\AA},
    $\beta=125.374^\circ$) and space group $C\,2/m$. The nearest neighbor Rh-Rh and Rh-O distances as well as Rh-O-Rh
angles are given. Note that the hexagon structure is not perfect and
there are two Rh-Rh and three Rh-O nearest neighbours.}%
\label{table_structure}
\begin{center}%
\begin{tabular}
[c]{l|l|l|l|l}\hline
atom&position&$x$&$y$&$z$\\\hline
Rh& 4h & 0 & 0.333 & 1/2\\
Li& 2a & 0 & 0 & 0\\
Li& 4g & 0 & 0.660 & 0\\
Li& 2c & 0 & 0 & 1/2\\
O& 8j & 0.516 & 0.327 & 0.263\\
O& 4i & 0.002 & 1/2 & 0.7380\\\hline
\multicolumn{2}{c}{}&dist./angle 1&dist./angle 2&dist. 3
\\\hline
\multicolumn{2}{c}{Rh-Rh} & 2.951~{\AA} & 2.952~{\AA} & \\
\multicolumn{2}{c}{Rh-O} & 2.023~{\AA} & 2.030~{\AA} & 2.021~{\AA}\\
\multicolumn{2}{c}{Rh-O-Rh} & 93.2$^{\circ}$ & 93.8$^{\circ}$ & \\\hline
\end{tabular}
\end{center}
\end{table}

\begin{figure}[ptb]
\begin{center}
\includegraphics[width=0.8\columnwidth]{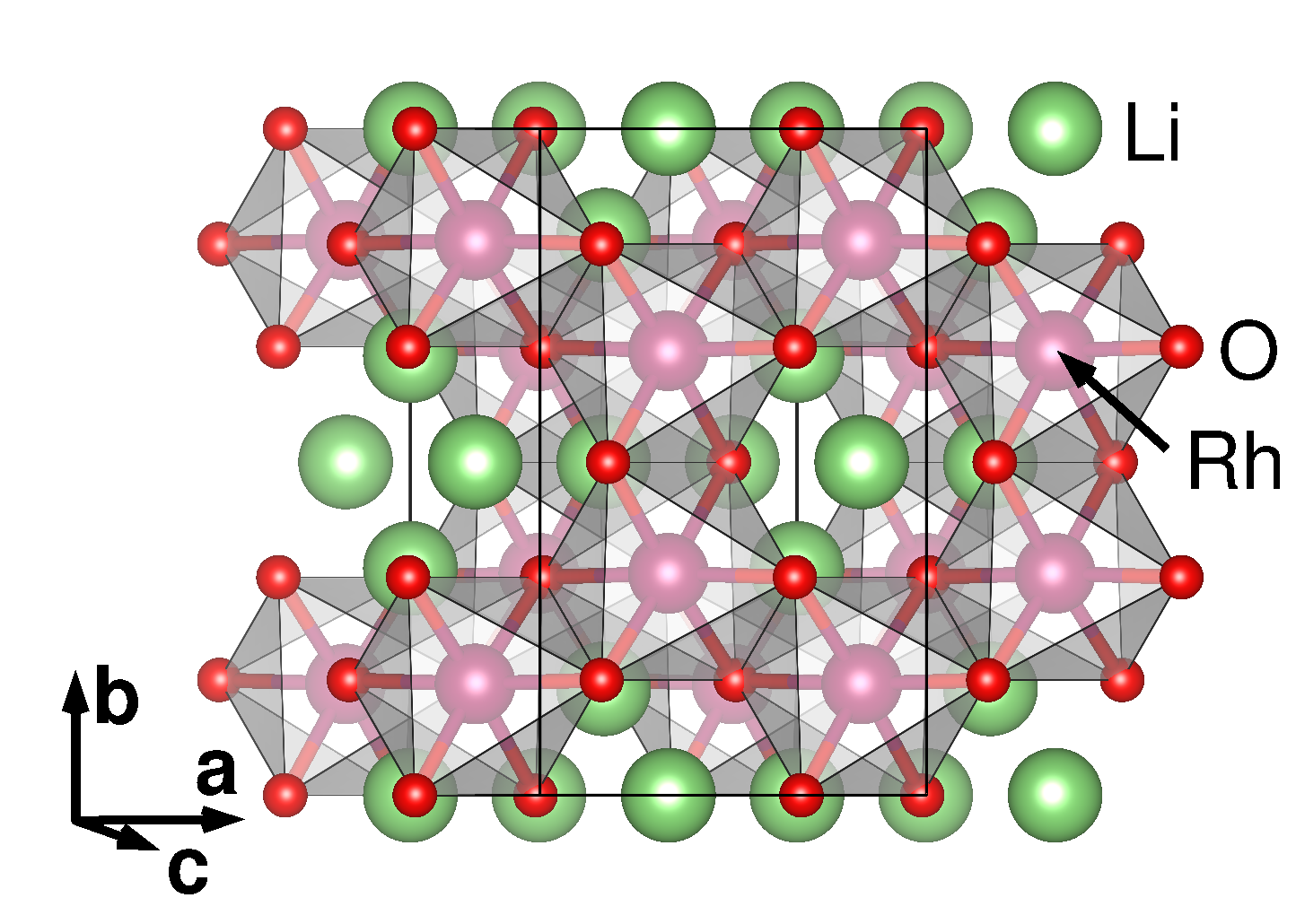}
\end{center}
\caption{(Color online) Structure of {\lirho}, viewed along $c^*$ direction.}
\label{fig:struct}
\end{figure}

For that reason, we have used (well-determined) unit-cell parameters
for {\lirho}, and performed first principles optimization of the
internal parameters.~\cite{Opt} We note the same procedure yielded
excellent agreement with the refined crystal structure of
{\nairo}.~\cite{Radu} The final structure is presented in
Table~\ref{table_structure} and shown in Fig.~\ref{fig:struct}. This
refined structure is consistent with the lab powder XRD data
(Fig.~\ref{powder}).


Despite the overall low crystal symmetry, the local symmetry of the
Rh$_{2}$Li planes is rather high: the hexagons are nearly ideal and
the Rh-O-Rh angles are nearly the same and relatively close to
90$^{\circ}.$ This makes it a showcase for the quasi-molecular orbital
concept.~\cite{Mazin12} To this end, we have performed first principle
calculations using the WIEN2k code,~\cite{Wien2k} and projected the
results using a standard Wannier function projection technique as
proposed by Aichhorn {\it et al.}~\cite{Aichhorn2009} and further developed in
Ref.~\onlinecite{FFJV}.  The resulting tight-binding parameters are shown in
Table~\ref{TB}.

\begin{table}[ptb]
  \caption{Comparison of the first principles hopping amplitudes in 
    {\lirho} and {\nairo}. The notations are explained in detail
    in Ref.~\protect\onlinecite{Foyevtsova13}. All hoppings $t$ and onsite energies $\mu$ are given in meV.}
\label{TB}
\begin{center}
\begin{tabular}{lcccc}
\hline\hline
& \multicolumn{2}{c}{Na$_{2}$IrO$_{3}$} & \multicolumn{2}{c}{Li$_{2}$RhO$_{3}$} \\
\hline
$\mu^{xy}$    & \multicolumn{2}{c}{-448.8} &\multicolumn{2}{c}{-385.8}\\
$\mu^{xz,yz}$ & \multicolumn{2}{c}{-421.5} &\multicolumn{2}{c}{-385.7}\\
$t_0^{xy\to xz,yz}$ & \multicolumn{2}{c}{-27.8} & \multicolumn{2}{c}{-18.8}\\
$t_0^{xz\to yz}$ & \multicolumn{2}{c}{-23.1} & \multicolumn{2}{c}{-15.5}\\
\hline
Distance&  3.130\,\AA & 3.138\,\AA &
2.951~\AA & 2.952~\AA \\
$t_{1\,{\rm O}}$ &269.6 &264.4 & 211.8 & 197.5\\
$t_{1\sigma}$    &-20.7 & 25.4 & -89.0 & -106.4 \\
$t_{1\perp}$$^*$     &-25.6/-21.4 & -11.9 & -15.9/-10.5 & -13.0 \\
$t_{1\parallel}$$^{\dagger}$ & 47.7/30.0&33.1 & 58.3/57.2 & 60.4\\
\hline 
Distance&  5.425\,\AA & 5.427\,\AA &
5.088~\AA & 5.096~\AA \\
$t_{2\,{\rm O}}$ & -75.8 & -77.0 & -77.2 & -78.7\\
$t_{2a}$$^{\dagger}$ & -3.5/-0.6 & -1.4 & -4.4/-5.3 & -4.3\\
$t_{2b}$ & -1.5 & -1.4 & 0.1 & 1.4\\
$t_{2c}$ & -36.5 & -30.4 & -24.9 & -24.1\\
$t_{2d}$$^*$ & 12.5/10.2 & 9.3 & 18.4/17.9 & 18.7\\
$t_{2e}$$^*$ & -21.4/-18.6 & -19.0 & -7.4/-7.8 & -7.6\\
\hline\hline
\end{tabular}\\
{\footnotesize $^*$For the shorter distance, the first number corresponds to $xy\to xz$, $xy\to yz$ transitions
  and the second to $xz\to yz$ transitions.\\
  $^{\dagger}$For the shorter distance, the first number corresponds to $xy\to xy$ transitions
  and the second number to $xz\to xz$, $yz\to yz$ transitions.}
\end{center}
\end{table}

\begin{figure}[ptb]
\begin{center}
\includegraphics[width=1.0\columnwidth,angle=0]{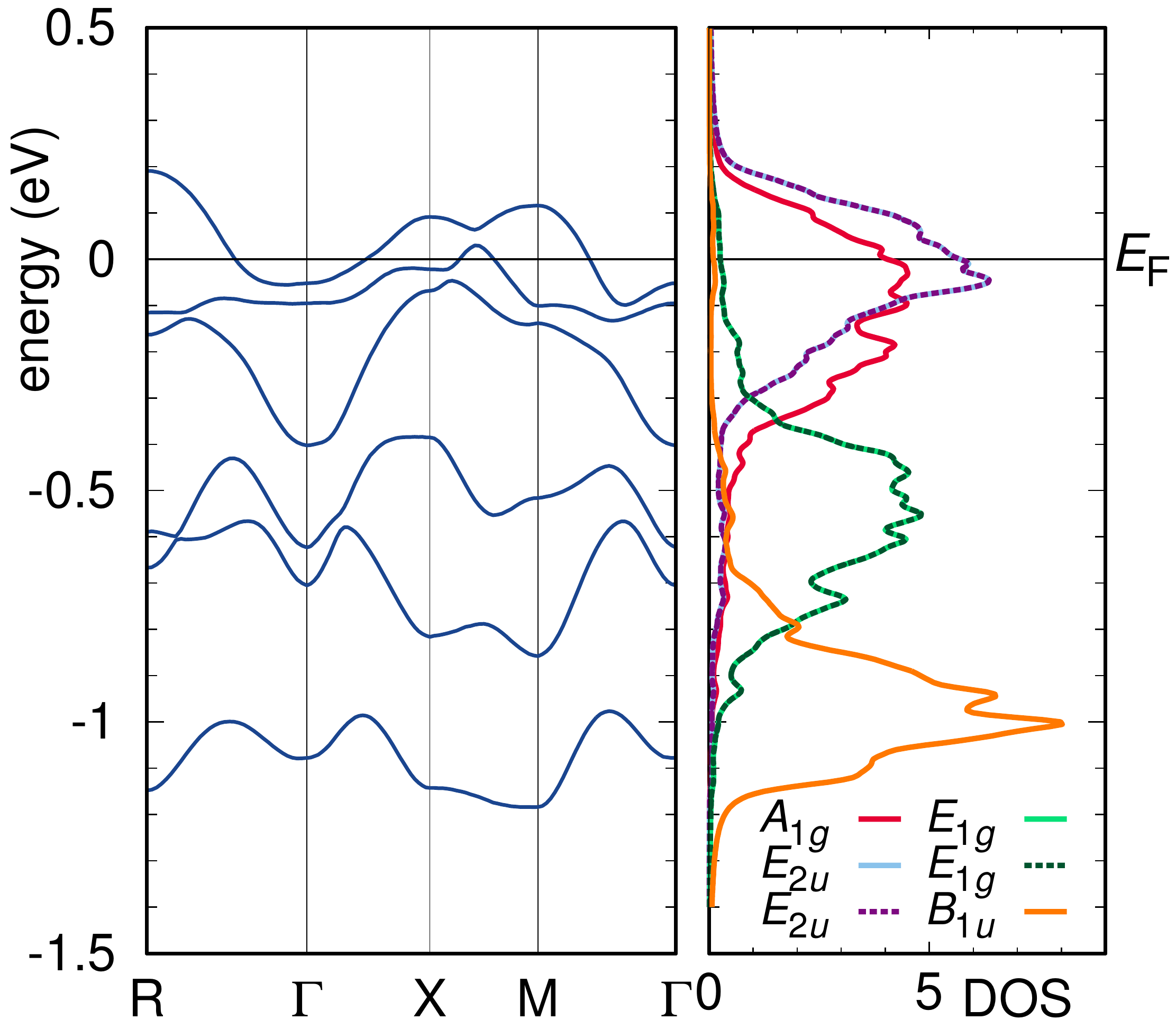}
\end{center}
\caption{(Color online) Nonrelativistic nonmagnetic band structure
  and density of states of {\lirho}, projected onto quasi-molecular
  orbitals, as described in Ref.~\onlinecite{Foyevtsova13}.}%
\label{DOS_nonrel}%
\end{figure}

As we see, the main condition for the QMO picture (dominance of the
O-assisted nearest neighbor hoppings) is fulfilled.  Projecting the
density of states (DOS) onto individual QMOs we see that, although it
does not separate into isolated manifolds as in {\nairo}, it is
composed of overlapping QMOs as shown in Fig.~\ref{DOS_nonrel}.

\begin{figure}[tbh]
\begin{center}
\includegraphics[width=0.47\textwidth]{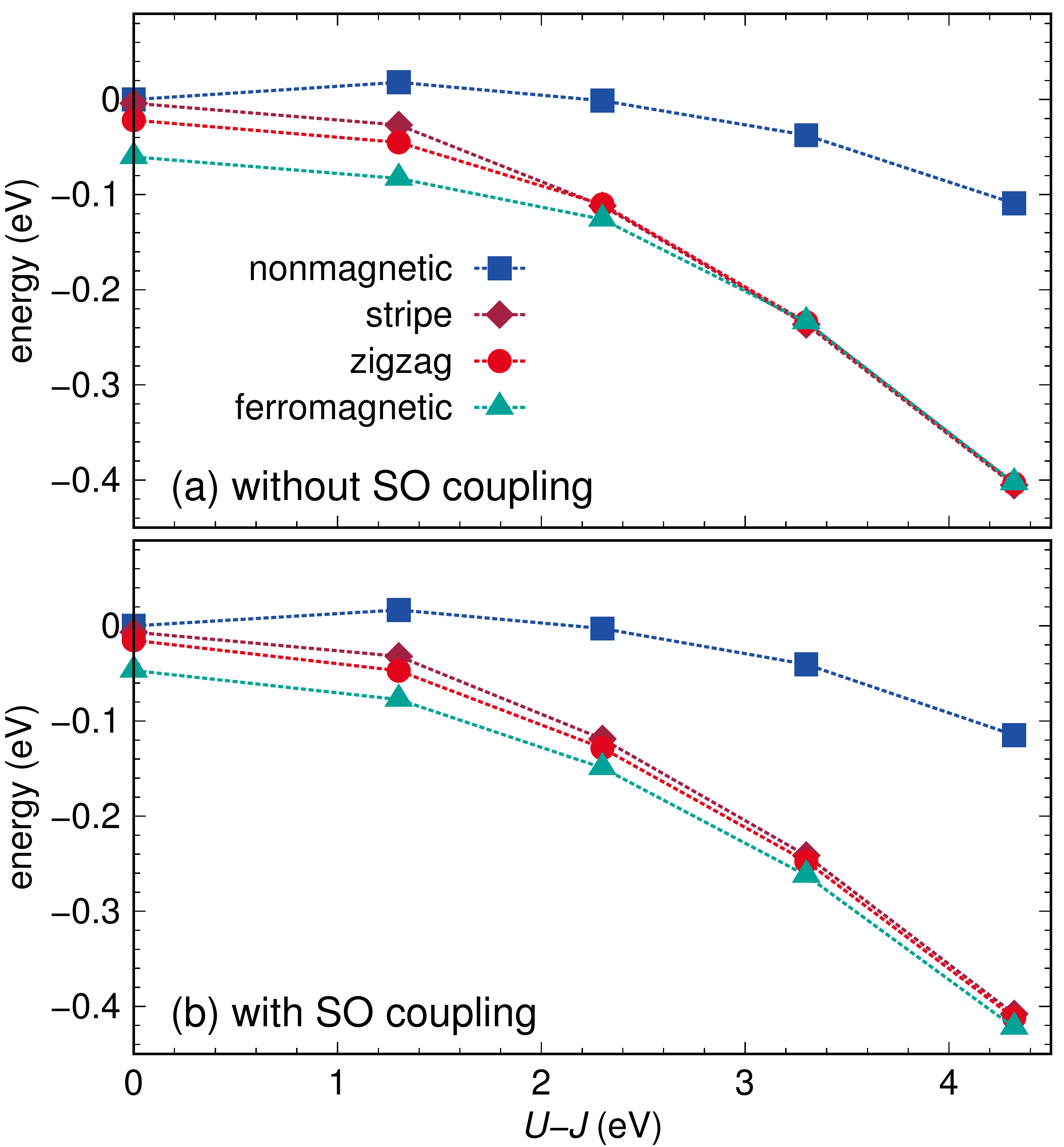}
\end{center}
\caption{(Color online) Energy of different magnetic configurations in
  eV/Rh relative to the nonmagnetic state, as a function of
  $(U-J)$. Energies at $(U-J) \neq 0$ are offset by $1.05(U-J)$. The
  top panel (a) does not include spin-orbit coupling, the bottom panel
  (b) does.}
\label{E}
\end{figure}

We have also performed spin-polarized calculations with various spin
configurations~\cite{Note} (Fig.~\ref{E}~(a)). We were not able to
stabilize a N\'{e}el order (magnetic moments collapse), but the
ferromagnetic (FM) and two antiferromagnetic phases, the
{\textquotedblleft}stripy{\textquotedblright} and the
{\textquotedblleft}zigzag{\textquotedblright} phase, are all stable,
{with the ground state practically degenerate between the two AFM
  states}. The FM state has a small advantage in the calculations,
which is lost upon application of $U$ (see below). 
The calculated FM state, just as in {\nairo}, is a half-metal with
$M=1\mu_{B}$/Rh. One has to keep in
mind that at small $U$ the material is metallic, which promotes 
ferromagnetism, and that LDA/GGA include spurious Hund's rule self-coupling of 
an orbital with itself. In particular, for the $\approx 90^o$
geometry, as in this case, the Hund's rule coupling on oxygen 
is not supposed to promote ferromagnetism~\cite{Chaloupka13,Foyevtsova13}
but in LDA/GGA it gives additional energetical
advantage to the ferromagnetic state of the order of $3I_O m_{\rm O}^2/4
\approx 3\cdot 1.6$ eV $\times 0.1^2/4\approx 12$ meV per Fe, where $I_{\rm O}=1.6$
eV is the Stoner factor~\cite{Mazin1997}, $m_{\rm O}=0.1$ the calculated magnetic moment of O, and there are three
oxygens per Fe.

\begin{figure}[ptb]
\begin{center}
\includegraphics[width=1.0\columnwidth]{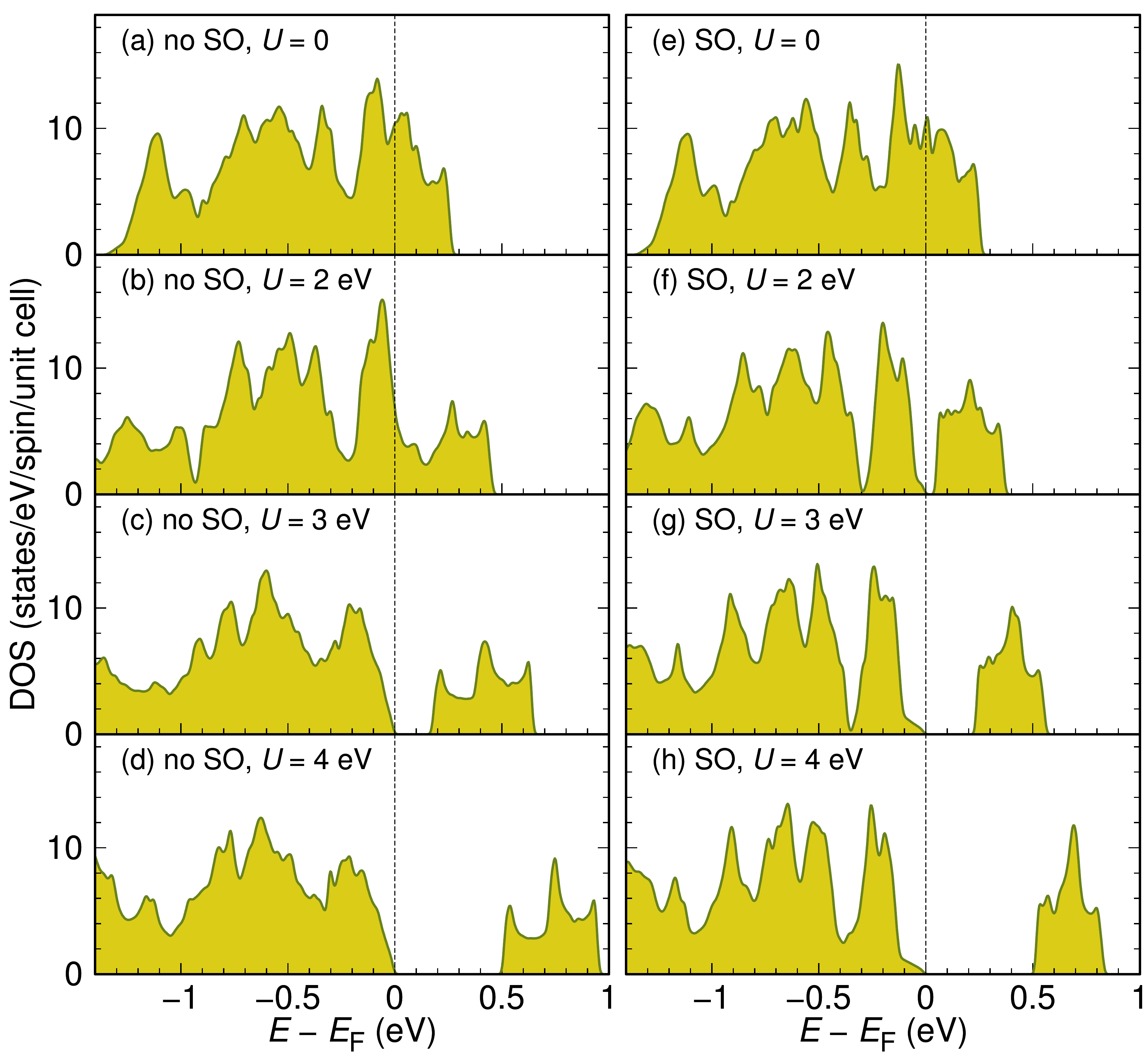}
\end{center}
\caption{(Color online) Evolution of the density of states with the
  Hubbard $U$ in the zigzag phase without spin orbit coupling (left
  panels) and with spin orbit coupling (right panels). We use
  $J_H=0.7$~eV throughout. }
\label{alldoszz}
\end{figure}

\begin{figure}[ptb]
\begin{center}
\includegraphics[width=0.48\textwidth]{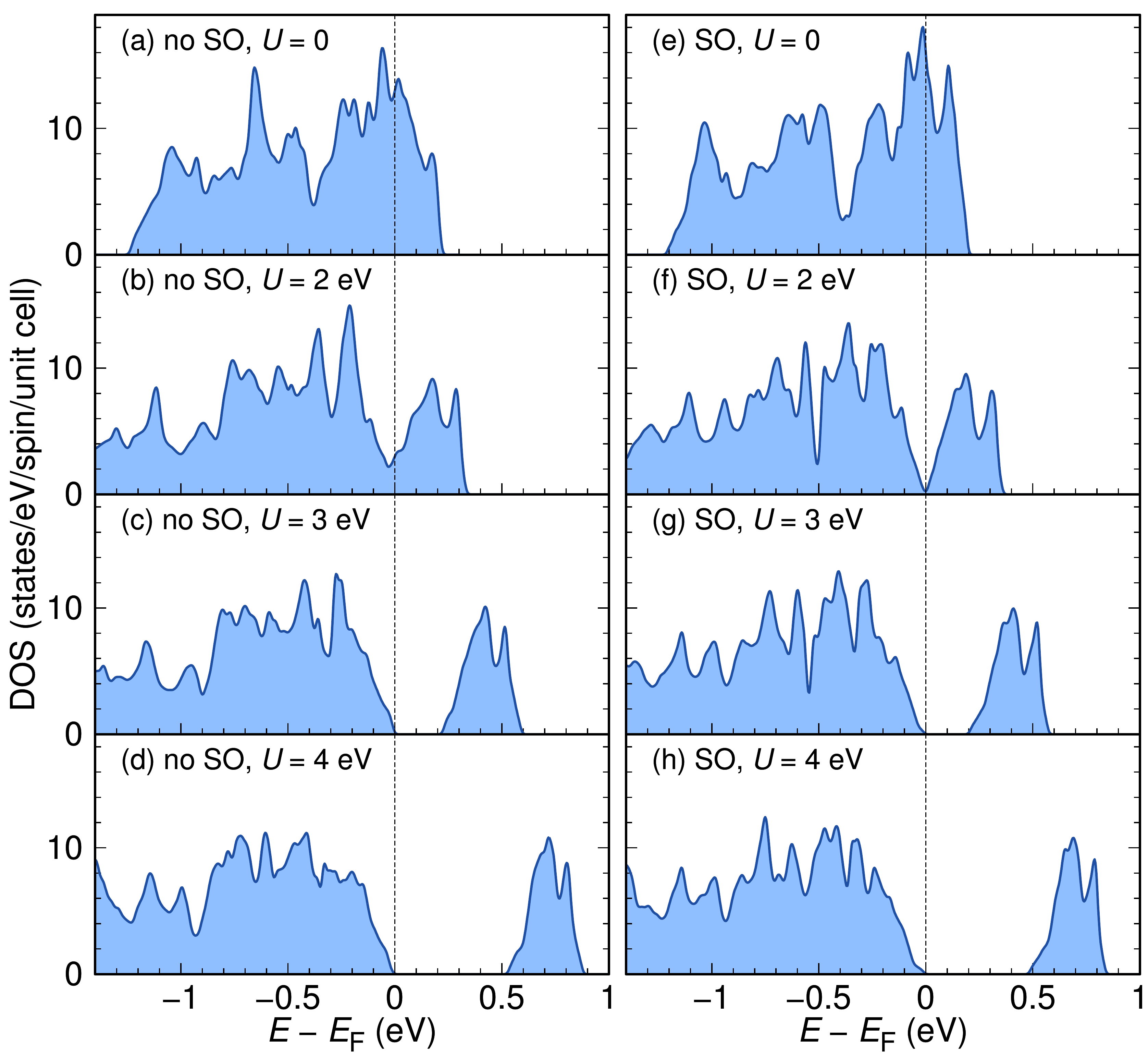}
\end{center}
\caption{(Color online) Evolution of the density of states with the
  Hubbard $U$ in the stripy phase, without spin orbit coupling (left
  panels), and with spin orbit coupling (right panels).  We use
  $J_H=0.7$~eV throughout. }
\label{alldosst}
\end{figure}

Neither the nonmagnetic state (Fig.~\ref{DOS_nonrel}) nor any of the
magnetic states considered (FM, stripy and zigzag) are
insulating. Including SOC has little effect on either energetics or
proximity to an insulator (Figs.~\ref{E}, \ref{alldoszz} and
\ref{alldosst}).

On the other hand, experimentally this material appears to be
insulating. It is natural to attribute this fact to the effect of
Hubbard $U$, which in 4$d$ metals is about 3-4 eV, twice as large as
for 5$d$ systems. Even though LDA+(on-site)U is a rather naive way to
tackle correlations in a QMO system, we have tried, \textit{faut de
  mieux}, to apply a standard LDA+U correction to our
calculations.~\cite{LDAU} As expected, for $U\gtrsim3$ eV we obtain an
insulator for both AFM configurations. In Figs.~\ref{alldoszz} and
\ref{alldosst} we show the evolution of the density of states (DOS)
with the Hubbard $U$ for the zigzag and stripy configurations,
respectively.  Besides, adding $U$ produces somewhat less obvious
effects. First, it destabilizes the FM structure, making all three
magnetic structures degenerate within the computational accuracy (in
calculations with SOC the FM state is a few meV lower in energy, but,
as mentioned, DFT always slightly overestimates this tendency because
it includes Hund's rule self-interaction on O). Second, the Hubbard $U$ 
enhances the SOC, increasing the calculated orbital moments.
The spin moment also positively correlates with $U$, but the dependence is much weaker. 
For instance, for the FM state the moment inside the Rh muffin
tin sphere increases from 0.58 to $0.66 \mu_{B}$ as $U$ increases from
0 to 5 eV.

An important point to make is that, as one can expect from the small
value of the SOC, it is not essential for obtaining an insulating
state; an antiferromagnetic order, however, is, just as in such
prototype Mott insulators as FeO and CoO. In fact, sometimes in LDA+U calculations including SOC is necessary
for reproducing the insulating behavior, even though a material is
obviously not relativistic. This is an artifact resulting from the
inability of LDA+U to describe  Mott insulators in
the paramagnetic case. One of the pathologies that LDA+U shares with
LDA is absence of local magnetic fluctuations~\cite{Ortenzi}.  In both
methods instead of a {\it para}magnetic state, {\it i.\,e.} a state
with disordered local moments, a {\it non-}magnetic state, with no
moments at all is considered. As a result, in such prototype strongly
correlated materials as for instance 3d oxides a metallic state is
protected by symmetry, unless some magnetic ordering is included (in
some cases even the ferromagnetic order suffices, in others an
antiferromagnetic ordering is needed), and LDA+U fails to reproduce
the paramagnetic insulating phase.  {\lirho} is a similar case.  In
the nonmagnetic calculations there are band crossings protected by
symmetry, that is to say, the best one can possibly achieve within
LDA+U, even with an arbitrarily large $U$, is a zero-gap
semiconductor. SOC, even infinitesimally small, removes this
protection (the protected bands can now hybridize), and now a
sufficiently large $U$ can open a full gap. Obviously, this fact does
not tell us anything about the real role of the SOC, but only
highlights shortcomings of the LDA+U method.~\cite{cao} In fact, while
the physics of {\nairo} and {\lirho} compounds is similar, the role of
interactions is reversed. In the former, strong SOC renders the
material nearly insulating already in the paramagnetic phase, and the
relatively weak correlations only help the existing tendency. In the latter, for $U=0$ there exists already a sizable separation
between the upper two bands but they are too wide and still overlap,
forming a negative gap (see Fig.~\ref{DOS_nonrel}). Indeed, strong correlations are
essential to open an actual gap, and the way to take correlations into
account in LDA+U is to include magnetism from the very beginning. This
scenario with the preexistence of a band separation is in contrast to
the case of Mott insulators and corresponds to a correlated band
insulator. 

It is worth noting that the ``213'' honeycomb structure is peculiar in
the sense that in the nearest neighbor approximation the highest band
is always a doublet, independent of the relative strength of the
SOC. In the strong SOC limit this doublet is the relativistic $j_{\rm
  eff}=1/2$. In the opposite limit, this doublet is an $A_{1g}$ molecular orbital, and
the band structure can be characterized as incipient band
insulator. Mott-Hubbard correlations obviously enhance the tendency to
insulating behavior, but generally speaking are not always
necessary. In real materials beyond this approximation the order of
states may change, in which case SOC becomes absolutely essential
(cf. {\nairo}), or band width may become too large for such a
simplistic treatment, but the fact that the most basic model has this
unique feature is very important for understanding the physics of
these honeycomb compounds. For a more detailed discussion we refer the
reader to Ref.~\onlinecite{Foyevtsova13}.

The observation that three different magnetic configurations, FM, zigzag and
stripy, with ordered moments on Rh ($\sim$0.5-0.7 $\mu_{B})$
independent of the magnetic pattern, are very close in energy
indicates considerable frustration.  Structural disorder is then
expected to push the system towards a spin-glass regime.

These results show an important similarity between the $5d$ compound
{\nairo} and the isostructural and isoelectronic $4d$ compound
{\lirho}, despite a much larger Hubbard $U$ and much smaller
spin-orbit $\lambda$ in the latter. This similarity suggests that
properties of these materials are largely controlled by the
non-relativistic, one-electron physics, namely the formation of
quasi-molecular orbitals, while the role of Coulomb correlations and
SOC lies primarily in enhancing already existing tendencies (in
particular, toward insulating behavior). As a word of caution, we want
to emphasize that while our results point toward these systems being
band (Slater) insulators rather than Mott insulators, this does not
indicate that they are weakly correlated or that they are localized
rather than itinerant.  On the other hand, our results suggest that
local antiferromagnetism is an important ingredient in the formation
of an insulating state and that Coulomb correlations are instrumental
in enhancing the insulating gap.

We thank Yogesh Singh and Radu Coldea for collaboration and valuable
discussions.  R.V. and H.O.J. acknowledge support by the DFG through
grants SFB/TR 49 and FOR 1346. Work in G\"ottingen is supported by the
Helmholtz Association through project VI-521. S.M. acknowledges
support from the Erasmus Mundus Eurindia Project. I.I.M. acknowledges
funding from the Office of Naval Research (ONR) through the Naval
Research Laboratory's Basic Research Program, and from the Alexander
von Humboldt Foundation.

 $^*$ Present address: Oak Ridge National Laboratory, P.O. Box 2008
Oak Ridge, TN 37831-6114


\begin{thebibliography}{99}
\bibitem{Opt} For initial optimization we used the VASP
  method~\cite{VASP} with the Perdew-Burke-Ernzerhof generalized
  gradient functional,~\cite{GGA} with the final adjustment using the
  all-electron WIEN2k code,~\cite{Wien2k} with default settings.  The
  latter code was also used for the total energy and DOS calculations.
\bibitem{Note} We double-checked the agreement of total energies
  between LAPW and FPLO.~\cite{FPLO}
\bibitem {LDAU} We used LDA+U with the so-called FLL (fully localized
  limit) double counting scheme, and with the Hund'd rule parameter
  $J=0.7$ eV.
\end{thebibliography}
\end{document}